# Termination control of NdGaO$_3$ crystal surfaces by selective chemical etching

V. Leca,[1,2,3,4] D. H. A. Blank[1], and G. Rijnders[1]

[1]Faculty of Science and Technology, MESA$^+$ Institute for Nanotechnology, University of Twente, PO Box 217, 7500 AE, Enschede, The Netherlands
[2]National Institute for Research and Development in Microtechnologies, Erou Iancu Nicolae Street 126 A, 077190, Bucharest, Romania
[3]Faculty of Applied Chemistry and Materials Science, Polytechnic University of Bucharest, Gheorghe Polizu Street 1-7, 011601, Bucharest, Romania

[4]Electronic mail: v.leca@alumnus.utwente.nl

**Abstract**
A chemical etching method was developed for (110) and (001) NdGaO$_3$ single crystal substrates in order to obtain an atomically flat GaO$_{2-\delta}$-terminated surface. Depending on the surface step density the substrates were etched in pH-controlled NH$_4$F- or NH$_4$Cl-based solutions, followed by an annealing step at temperatures of 800-1000$^o$C, in air or in oxygen flow, in order to recrystallize the surface. Atomic Force Microscopy (AFM) and high-pressure Reflection High Energy Electron Diffraction (RHEED) were used to analyse the surface morphology of the samples after every treatment. Studies on the chemistry and characteristics of the terminating layer showed that the chemically etched NdGaO$_3$ substrate surface has a GaO$_{2-\delta}$ termination and that the (110) and (001) NdGaO$_3$ surfaces are characterized by a different free surface energy, which is lower for latter.

## 1. Introduction

Studies of novel properties taking place at the interface between complex oxides, such as orbital reconstruction [1] or electronic coupling [2], requires fabrication of thin-film heterostructures with an atomic-scale roughness of the individual components. Development of methods for yielding substrates with unique and known composition of the terminating layer is of utmost importance for precise atomic control of the interfaces in such multilayer structures [2]. When growing films of compounds with layered (perovskite) structure, the composition of the substrate terminating layer determines the surface morphology of the film [3], as well as the stacking sequence [4], and therefore the final physical-chemical properties of the films [5, 6]. Substrates with single terminated surface are therefore required for reproducible thin film growth with respect to morphology and epitaxy, while the substrate-film interface quality will influence the physical properties of the films [6-9]. It is also desirable that the lattice mismatch between the substrate and the film is as small as possible in order for the film to grow following, preferably, a 2D growth mechanism [9].

NdGaO$_3$ single crystals, with either (001) or (110) orientation, belong to the class of substrates with layered structure and are often used as template for epitaxial growth of high critical temperature superconductors [9, 10], infinite-layer type structures (ACuO$_2$, with A=alkaline-earth metal cation) [11], colossal magnetoresistive materials [12], or other compounds (e.g., GaN, Sr$_2$RuO$_4$) [13, 14] due to the similarities in their crystal structure. NdGaO$_3$ is the only oxide among the lanthanide gallates with no structural phase transitions below $\sim$ 1000 $^o$C. Therefore, it is often used to grow twin-free thin films when deposition is done at a temperature lower than the phase transition temperature [15]. Low RF loss of NdGaO$_3$ (see table 1) makes this substrate more suitable for microwave applications than other substrates, such as

Table 1. Room temperature structural and physical properties of bulk $NdGaO_3$ [9, 16, 17]

| Symmetry | Space group | Unit cell (Å) | $m_p$ (°C) | Phase transf. (°C) | $\varepsilon$ | $\rho$ (g/cm$^3$) | Tan $\delta$ | $\alpha$ (K$^{-1}$) |
|---|---|---|---|---|---|---|---|---|
| Orthorhombic, GdFeO$_3$ type | Pbn2$_1$ | $a_o$ = 5.428 $b_o$ = 5.498 $c_o$ = 7.708 | 1605 | >1000 | 20-25 (1MHz) | 7.57 | $3\times10^{-3}$ | $5.8\times10^{-6}$ |

$\varepsilon$ = dielectric constant, tan $\delta$ = dielectric loss, $\alpha$ = thermal expansion coefficient, $m_p$ = melting point

SrTiO$_3$ [9]. X-ray [16], neutron diffraction [17], and synchrotron single crystal diffraction data [18] showed that NdGaO$_3$ has at room temperature a layered GdFeO$_3$ type structure (distorted perovskite with an orthorhombic symmetry corresponding to the space group Pbn2$_1$) [17] consisting of alternating stacks in the (001) direction of NdO$_{1+\delta}$ and GaO$_{2-\delta}$ atomic layers, with NdO$_{1+\delta}$ a basic oxide and GaO$_{2-\delta}$ an acidic oxide [17-19]. When the crystal is cut along the (110) plane, the resulted pseudo-cubic (pc) cell, which is slightly distorted in comparison with the primitive perovskite structure, have $a_{pc} = b_{pc} = 0.5(a_o^2+b_o^2)^{1/2}$ = 3.863 Å, $c_{pc} = c_o/2$ = 3.854 Å, $\gamma$ = 89.26°, where $a_o$, $b_o$, and $c_o$ are the cell parameters of the orthorhombic unit cell, given in table 1 [17].

The morphology of the commercial (as-received) substrates is determined by the polishing method used, their surface consisting of randomly distributed islands with typical height of (n + ½) unit cells resulting, therefore, in a mixed termination. Structural defects and impurities, such as C and organic compounds, are often present on the surface of the as-received substrates as a result of the polishing [20]. Low-temperature annealing (typically at 500-600 °C) in O$_2$ flow (ex-situ) or in vacuum is generally used for removal of the surface contaminants [21]. However, although the surface morphology after such thermal treatments is improved, step ledges are rather imperfect as shown in the AFM visualization of the surface morphology of an as-received (110) NdGaO$_3$ substrate (figure 1(a)), while the mixed character of the termination layer remains. The streaky RHEED pattern (recorded at 550 °C and $5\times10^{-2}$ mbar O$_2$) characteristic of such surface (figure 1(b)) indicates a flat, but disordered surface.

NdO$_{1+\delta}$ single terminated (001) NdGaO$_3$ substrate surface was obtained by annealing in air for 2h at high temperature (1000 °C) by Ohnishi *et al.* [22], the composition of the terminating layer being determined by coaxial impact-collision ion scattering spectroscopy. A similar result (a single terminated surface) has been reported for (110) and (001) NdGaO$_3$ substrates annealed in oxygen at temperatures of 800 – 950°C, but the composition of the termination layer was not specified [23, 24]. The results of annealing was shown to depend strongly on the substrate miscut angle, its surface orientation with crystallographic axes (defined here as the angle Δ), as well as on the ambient (vacuum, air, or O$_2$) used for annealing [22-26]. Mixed termination surfaces are typically observed if these parameters are not taken into account prior to the thermal treatment. Another reason for the difficulty in controlling the surface morphology of the annealed samples may come from the presence of polishing or post-polishing contamination on the surface [20, 21]. These impurities can be retained on the surface even after heat treatment at temperatures as high as 1100$^o$C, as was shown for SrTiO$_3$ [27], playing a crucial role in the evolution of the surface morphology. The best method of removing the surface contaminants, as well as for controlling the chemistry of the substrate termination layer, is considered the chemical etching [25-31]. Different chemical solutions (e.g., HCl, HNO$_3$, as well as a mixture of H$_2$SO$_4$:H$_3$PO$_4$ = 3:1) have been reported to be able of improving the surface morphology of the NdGaO$_3$ substrates, but no details on the influence of this process on the surface properties were given [13, 32]. Therefore, methods that can produce atomically flat single terminated NdGaO$_3$ substrates with known composition of the top layer and reproducible results, independent on the above-mentioned substrates characteristics, are still required.

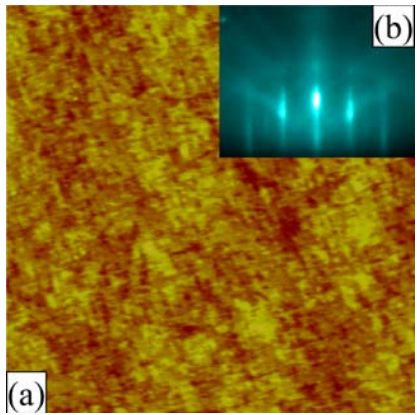

**Figure 1.** AFM image (a) and the corresponding RHEED pattern (b) of an as-received (110) NdGaO$_3$ substrate annealed for 0.5 h at 550 °C in O$_2$ flow (100 sccm). No terrace structure can be observed. AFM scan size: $4\times4$ μm$^2$.

In this paper we present the results of a chemical etching procedure developed for yielding structurally and chemically well-defined surfaces for (110) and (001) NdGaO$_3$ single crystal substrates. The etching method considers the layered structure of the substrate material and the difference in chemical reactivity of the constituent oxides [19]; one of the surface oxides (i.e., the basic oxide, NdO$_{1+\delta}$) is selectively removed by means of a chemical reaction, yielding a GaO$_{2-\delta}$ terminated surface [28, 31]. In finding the right etching conditions (i.e., the etching time, the pH of the etchant), one has to consider several parameters, such as: the characteristic surface properties of each substrate (e.g., the type of constituent oxides, the miscut angle and its orientation angle, $\Delta$), the correlation between etching speed and the miscut angle (the etching being more aggressive with increased surface step density) [30], and the reactivity (pH) of the etching solution. An increased value of the miscut angle implies an increased probability of kink and ridge sites on the surface. As the etching removes material predominantly from the sides of these sites (the kink sites being etched more easily than the ridge sites) [30], the reactivity of the chemical solution has to be adapted to the substrate miscut angle, $\alpha$. Therefore, since the etching process depends on the local atomic-scale configuration, an increased step density at the substrate surface is expected to result in an increased etching speed.

Following the above considerations, two types of chemical solutions were developed and used during this work: a modified commercial buffered-HF solution (HF + NH$_4$F + H$_2$O, named here m-BHF), and a buffered-HCl solution (HCl + NH$_3$ + H$_2$O, named here BHCl). The latter is developed for wet etching of NdGaO$_3$ substrates characterized by $\alpha > 0.3^o$. The HCl (35 vol%), NH$_3$ (25 vol%), and the commercial buffered HF (BHF, NH$_4$F:HF=87.5:12.5) solutions used for the experiments were produced by Merck. The influence of the type of NdGaO$_3$ substrate and their surface morphology on the etching characteristics using these NH$_4$F- and NH$_4$Cl-based solutions will be presented and discussed. Wettability of TiO$_2$ and SrO (or BaO) on chemically and (only) thermally treated substrates is furthermore presented to obtain information on the chemistry of the terminating layer after such treatments. The surface morphology was analysed by means of high-pressure RHEED (in-situ) and by AFM (ex-situ). Both are surface sensitive techniques, RHEED giving a statistical view over a larger surface area, while AFM provides local information of the surface morphology. The RHEED patterns were recorded in a pulsed laser deposition (PLD) chamber equipped with a high-pressure RHEED system [33]. The patterns were recorded at $5\times10^{-2}$ mbar O$_2$ at a substrate heater temperature of 550 $^o$C. Topographical AFM images were acquired at room temperature in an atmospheric environment with a Digital Instruments Nanoscope III scanning probe microscope, in contact mode. Standard Si$_3$N$_4$ triangular contact mode tips ($k = 0.58$ Nm$^{-1}$) were used with the total interaction force kept as low as possible. Imaging took place after the substrates were treated (chemically and/or thermally) while the films (of $TiO_2$, SrO, or BaO, see next) were analyzed immediately after they were removed from the deposition chamber. This insures that surface contamination, that would affect the quality of the image, is minimized.

## 2. Experimental
### 2.1. Method description

The m-BHF etching solutions used for the studies presented here were prepared by considering the dissociation processes in the $HF + NH_4F + H_2O$ system. Due to the strong electrolyte properties of $NH_4F$, $F^-$ ions are generated in the $HF + NH_4F + H_2O$ solution as part of the ionisation process in the $HF + H_2O$ system [34]. A large amount of difluoride ions $HF_2^-$ (with higher reactivity than that of the $F^-$ ions) is then formed in solution. By decreasing the amount of $F^-$ and $HF_2^-$ ions through neutralisation with $NH_3$, the reactivity of the resulting buffered HF solution is reduced yielding the desired pH and, therefore, the necessary reactivity. Due to the difference in their specific free surface energy (as concluded from annealing experiments) [25], chemical treatments of (110) and (001) $NdGaO_3$ substrates were done in acidic solutions with different pH values, higher for former and lower for latter. Therefore, an etchant with a lower reactivity was used for removing the $NdO_{1+\delta}$ layer from the (110) $NdGaO_3$ surface. The composition of the m-BHF solution prepared for the (110) $NdGaO_3$ substrates consisted of 10 ml commercial BHF + 4 ml $NH_3$ (25 vol.%) + 90 ml de-ionized (DI) water, resulting in a solution with a pH of ~ 5.5. The DI water (pH=6-7, R=10-15 MΩ) was produced with Millipore Elix equipment. The m-BHF solution used for the (001) $NdGaO_3$ substrates consisted of 20 ml commercial BHF + 3 ml $NH_3$ (25 vol.%) + 110 ml DI water, with the resulting solution pH of ~ 4.5. In these m-BHF solutions the amount of $NH_3$ was determined so that it does not result in insufficient etching of the surface $NdO_{1+\delta}$.

The etching procedure consisted of several subsequent steps [31]. The substrates are first immersed in DI water for up to 20 min (depending on the miscut angle, the smaller this value the longer the soaking time), then in the m-BHF solution for 0.5-2 min, rinsed with DI water for about 20-30 seconds and then with ethanol, and finally dried in a nitrogen stream. All steps of the etching procedure are performed in an ultrasonic bath. To facilitate surface recrystallization a final (ex-situ) annealing step is performed at 800-1000 ºC for 0.5-4 h, in air or in $O_2$ flow (50-200 sccm). The kinetics involved in the surface reconstruction is strongly dependent on the thermal treatment parameters (temperature and time) [35]. These parameters were selected based on the substrate miscut angle, the lower this value, the higher the annealing temperature and the longer the annealing time.

Taking into account the differences in chemical properties of the constituent oxides of the $NdGaO_3$ substrate [22], the aim of the first step (soaking in DI water) is the selective transformation of one of the surface oxides, namely $NdO_{1+\delta}$, in a hydroxide complex, which can be easily removed by the etchant through transfer of the hydroxide complex into solution [19, 31]. In the first step improvement of the surface wettability is achieved and as a result the reactivity with the etchant is increased. The composition of the etching solutions was determined based on the reactivity of the $NdO_{1+\delta}$ and the solubility in water of the reaction product, $NdX_3 \cdot x(H_2O)$, where X = F, Cl (see next for HCl-based solutions). In order to remove any traces of the reaction product from the substrate surface, as the $NdX_3 \cdot x(H_2O)$ complex will decompose in the $NdO_{1+\delta}$ during the annealing step and subsequently crystallizing on the surface, the etched wafers were rinsed in DI water after they were immersed in the etchant.

Commercially available (110) or (001) $NdGaO_3$ single crystal wafers with dimensions of $10 \times 10 \times 1$ $mm^3$ were used for the annealing and chemical etching experiments described in this paper. The substrates, produced by Surfacenet GmbH (Rheine, Germany) and by MTI Corporation (Richmond, USA), were one side mechano-chemically polished, with the misorientation (the miscut angle) of the surface with respect to the (110) or (001) crystal plane of 0.02-0.5º.

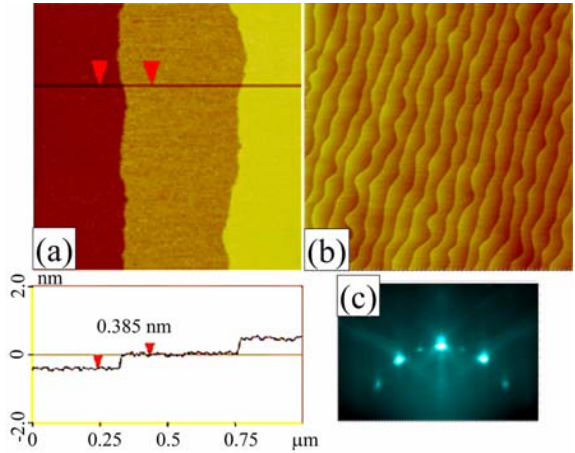

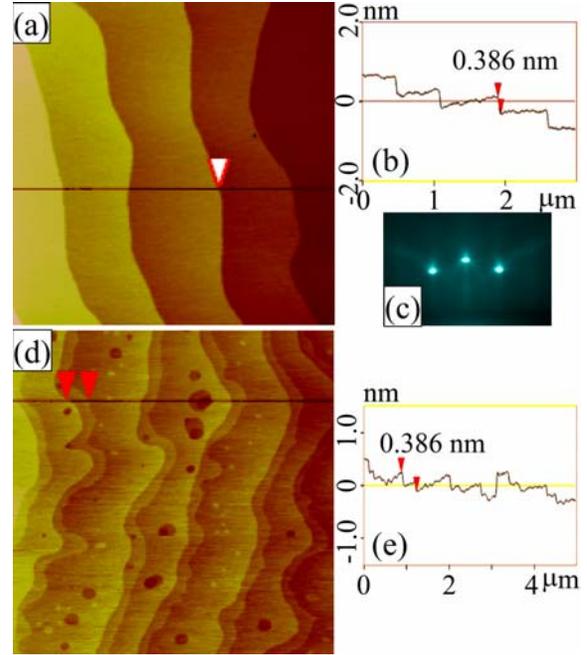

**Figure 2.** AFM images (a and b) and the corresponding RHEED pattern (c) of a (110) NdGaO$_3$ substrate chemically etched for 0.5 min in m-BHF solution (pH=5.5) and then annealed for 2 h at 1000 $^o$C in O$_2$ flow (100 sccm). The line scan profile from (a) shows one pseudo-cubic unit cell height terraces; the RHEED pattern reveals sharp Kikuchi lines which indicates an atomically smooth, crystalline surface. AFM scans size: (a) 1×1 μm$^2$, and (b) 7×7 μm$^2$.

The value of the miscut angle was determined by means of X-ray diffraction or from AFM data. Prior the annealing or chemical etching experiments the wafers were cut in smaller slices of about 5×5×1 mm$^3$ and then cleaned with chloroform (30 min), acetone (20 min), and ethanol (20 min) in order to remove the wax used during cutting and the resulted surface contamination. The ex-situ annealing experiments of the as-received or chemically etched substrates have been performed in a tube-oven, under flowing oxygen (at a rate of 50 to 200 sccm) or air. The substrates were placed on an Al$_2$O$_3$ boat inside the quartz tube of the oven. For re-growth or recrystallization of the step ledges the annealing temperature was selected above the threshold value (considered to be 600 $^o$C) [36]. The temperature was ramped to the desired value with a rate of 15 $^o$C/min. The annealing temperature, time and atmosphere used were selected taking into account the substrate orientation [(001) or (110)] and the value of the miscut angle.

**Figure 3.** Height AFM images (a and d) and line scan profiles (b and e) of (001) NdGaO$_3$ substrates chemically etched for 2 min in m-BHF solutions with (a) pH=4.5, and (d) pH=5.5, followed by annealing for 2 h at 1000 $^o$C in air or O$_2$ flow (200 sccm). The 2D RHEED pattern from (c) indicates and atomically flat surface for the substrate from (a). AFM scans size: (a) 3×3 μm$^2$, and (d) 5×5 μm$^2$.

### 2.2. Experimental results and discussion

Topographic AFM micrographs and the corresponding RHEED pattern of an etched (110) NdGaO$_3$ substrate (miscut angle of ~0.05$^o$) are shown in figure 2. The wafer was chemically etched in m-BHF solution (pH=5.5) for 1 min and then annealed for 2 h under O$_2$ flow (100 sccm) at 1000$^o$C. A surface morphology characterised by a terraced structure with steps of one pseudo-cubic unit cell height has resulted, free of step bunching or etched pits, as shown by the topographic AFM image and by the line scan profile (figure 2(a)). A larger area view of the surface morphology of the same substrate is given in figure 2(b). The corresponding RHEED pattern (figure 2(c)) shows sharp and narrow diffraction spots, as well as Kikuchi lines characteristic of a surface with high crystallinity. Similar results yielded the etching of (001) NdGaO$_3$ substrates in m-BHF solution (pH = 4.5), followed by annealing at temperatures of 950-1000 $^o$C in air or in O$_2$, as shown by the AFM data (figure 3(a) and the corresponding line scan profile from figure 3(b)) and by the RHEED pattern (figure 3(c)) for a substrate chemically etched for 2 min and annealed for 2 h at 1000 $^o$C in O$_2$ flow (200 sccm). The terraces height in figure 3(a) corresponds to the distance between two consecutive crystallographic planes with the same composition, i.e. c$_o$/2.

An etching time of 0.5-2 min was found to be optimum for substrates with miscut angle value <0.2$^o$, independent of the substrate orientation. However, etching of the (001) NdGaO$_3$ substrates with the m-BHF solution (pH=5.5) developed for the (110) oriented substrates resulted in an incomplete etching, the AFM image of such surface showing a mixed termination, as can be observed in figures 3(d) and 3(e). A longer etching time resulted in a decrease of the amount of NdO$_{1+\delta}$ on the surface, but also in the presence of etched pits in the GaO$_{2-\delta}$ layer, while the mixed termination character of the surface did not change. This result confirmed the difference in the free surface energy for the (110) and (001) NdGaO$_3$ substrates, which is lower for latter, as was also found from the annealing experiments [28, 29].

Lowndes *et al.* [37] showed that if the substrate miscut angle is sufficiently large spiral and island growth can be completely eliminated for YBa$_2$Cu$_3$O$_{7-x}$ thin films deposited on a nearly lattice matched substrate (e.g., SrTiO$_3$, NdGaO$_3$). Since a surface with an increased step density is characterized by a higher density of kink and ridge sites, a buffered-HCl solution, with chemical reactivity lower than that of the m-BHF solutions was developed. This BHCl solution, with a pH of ~ 1.5, consisted of 10 ml HCl (35 vol.%) + 4 ml NH$_3$ (25 vol. %) + 90 ml DI water. The role of NH$_4$OH is to maximize the solubility of the etching reaction product (NdCl$_3$·$x$H$_2$O) which is easily soluble in NH$_4$OH [19], local saturation of the reaction product resulting in slowing or suppression of the etching process [34]. The AFM image of a (001) NdGaO$_3$ wafer (with miscut angle α~0.30°) etched in BHCl solution for 5 min and annealed for 1 h in O$_2$ flow (200 sccm) at 950 $^o$C is given in figure 4(a). It is important to mention that with increasing miscut angle annealing requires moderate temperatures (the higher the miscut angle value, the lower the annealing temperature and the shorter the annealing time), otherwise thermal roughening may result, due to segregation effects that may introduce Nd residue on the surface, as well as step bunching formation. Step bunching was also observed in case of the NdGaO$_3$ substrates with α > 0.3 treated in BHF-based solutions. For comparison, in figure 4(b) is shown the surface morphology of a (001) NdGaO$_3$ wafer (characterised by the same miscut angle as the substrate from figure 4(a)) etched for 0.5 min in m-BHF solution (pH = 4.5) and annealed under the same conditions. A terraced structure with steps of one-unit cell (c$_{pc}$) height was observed for both substrates from figure 4, with step bunching presented for the wafer etched in the m-BHF solution (figure 4(b)). Similar results were also obtained when a commercial BHF solution was used instead of the m-BHF.

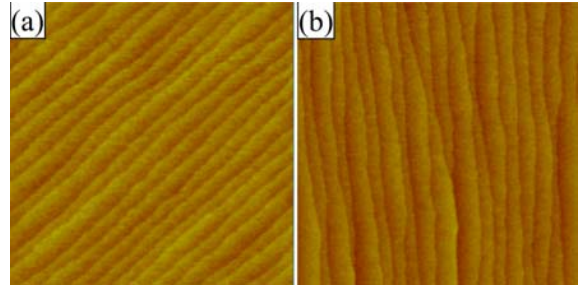

**Figure 4.** AFM images of (001) NdGaO$_3$ substrates chemically etched and annealed 1 h at 950 $^o$C in O$_2$ flow (200 sccm): (a) etched for 5 min in BHCl solution, and (b) etched for 0.5 min in m-BHF solution. Scans size: 1×1 μm$^2$.

The composition of the topmost layer of the etched substrates was determined by taking into account that the NdO$_{1+\delta}$ and GaO$_{2-\delta}$ surfaces are chemically different. Since the annealing of the as-received NdGaO$_3$ substrates results in a NdO$_{1+\delta}$ termination [22], the specific surface termination for the chemically treated NdGaO$_3$ substrates may be determined from: i) the difference in surface morphology after chemical and (only) annealing treatments, ii) the difference in wetting of, e.g., TiO$_2$ and SrO (or BaO) on chemically etched and on (only) annealed substrates, and iii) from the stability in air of the chemically etched and of the (only) annealed surfaces. The results of each procedure will be presented next (where chemical treatment implies wet etching followed by an annealing step).

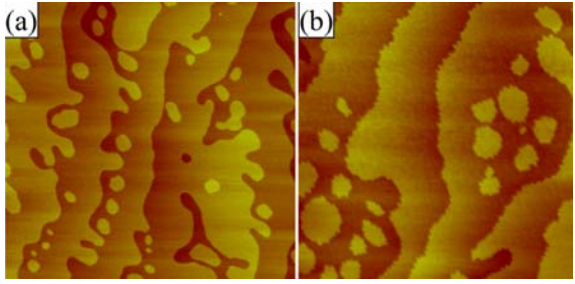

**Figure 5.** AFM images of (001) NdGaO$_3$ substrates (cut from the same larger wafer) characterized by an angle Δ of 45°: (a) chemically etched for 2 min in m-BHF solution and annealed for 2 h at 1000 °C in air, (b) only annealed for 2 h at 1000 °C in air. Scans size: (a) 7×7 μm$^2$, and (b) 5×5 μm$^2$.

  i. Two slices were cut from one (001) NdGaO$_3$ wafer (characterized by an angle Δ ~ 45°) and then cleaned with chloroform, acetone and ethanol, as described above. Next, one slice was chemically etched in m-BHF solution for 2 min, but without annealing it. The resulting etched slice and the as-received one were then annealed simultaneously for 2 h at 1000°C, in air. Contact-mode AFM micrographs (figure 5) reveal surfaces characterized by terraces with two types of edges. For the chemically etched sample (figure 5(a)) the terraces edges have a rounded (curved) appearance, i.e., the edge energy is minimized. This is in contrast with the morphology of the annealed surface (figure 5(b)), which shows straight and sharp

Table 2. Deposition parameters for the MeO$_x$ (Me=Ti, Sr, Ba) thin films.

| Material | T$_d$ (°C) | P$_d$ (Pa) | E$_d$ (J/cm$^2$) | f (Hz) | d$_{st}$ (mm) |
|---|---|---|---|---|---|
| TiO$_2$ | 850 | 7.5 | 2 | 1 | 45 |
| SrO | 825 | 10 | 2 | 1 | 45 |
| BaO | 590 | 15 | 1.3 | 1 | 45 |

T$_d$ = deposition temperature, P$_d$ = O$_2$ deposition pressure, E$_d$ = laser energy density on target, f = laser repetition rate, d$_{ts}$ = substrate-target distance edge step edges (saw-tooth like) aligned along the [100] and [010] crystallographic directions. Hence, it may be concluded that the surface of the chemically etched substrate is characterized by a lower surface energy and has a different chemical composition than that of the annealed one. Similar differences in the type of terraces edges and chemistry of the termination layer were observed for annealed and chemically etched (001) SrTiO$_3$ substrates [3, 38].

  ii. The wetting of NdO$_{1+\delta}$ and (presumably) GaO$_{2-\delta}$ terminated surfaces was determined by deposition of one unit cell thick TiO$_2$, SrO or BaO layers by means of PLD on either annealed or chemically treated substrates. The amount of deposited material (14 laser shots for TiO$_2$ and SrO, and 12 laser shots for BaO, respectively) was less than (but close to) one monolayer in order to avoid formation of precipitates [39]. Deposition conditions for each oxide are given in table 2. All targets were single crystals. First, TiO$_2$ was deposited on an annealed (110) NdGaO$_3$ surface. Figure 6(a) shows the AFM and RHEED data of the resulting film surface. Regularly spaced terraces separated by steps of ~0.4 nm height are visible. It can be concluded that TiO$_2$ is wetting the NdO$_{1+\delta}$ on the surface forming an almost closed layer. The corresponding RHEED pattern indicates a highly ordered surface. Since a B-site oxide (e.g., TiO$_2$) does not wet a B-site terminated ABO$_3$ surface [3, 39, 40] (e.g., a GaO$_{2-\delta}$ terminated NdGaO$_3$), it can be concluded that the annealed NdGaO$_3$ has a NdO$_{1+\delta}$ termination, confirming the results from Ref. 22. SrO was next deposited on a m-BHF etched (001) NdGaO$_3$ surface. The RHEED specular intensity exhibited clear oscillatory behavior, the deposition being stopped close to completion of one monolayer. The resulting surface morphology and the corresponding RHEED pattern of the film are shown in figure 6(b). The AFM micrograph shows a smooth surface with unit cell steps, while the RHEED pattern is characterized by sharp Bragg diffraction dots indicating a well-defined surface. The wetting is complete, giving rise to an almost closed layer. Similar results were obtained for BaO (instead of SrO) deposited on chemically treated NdGaO$_3$ substrates, with either (110) or (001) orientation, suggesting that the chemically and the (only) thermally treated NdGaO$_3$ substrates are characterized by different terminations.

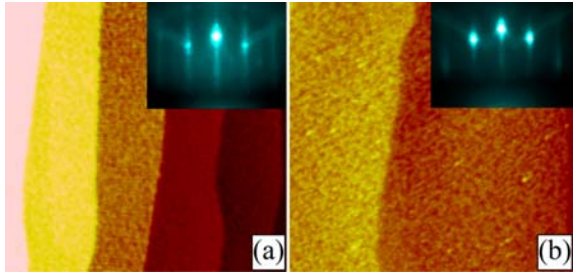
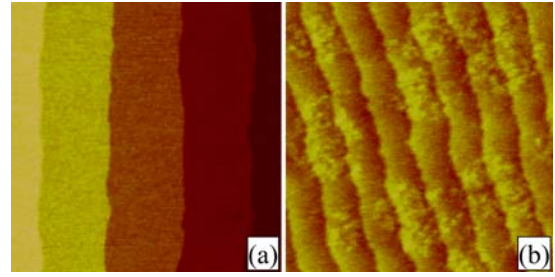

**Figure 6.** (a) AFM image (1.5×1.5 μm$^2$) and the corresponding RHEED pattern of ~ 1 monolayer TiO$_2$ deposited on an annealed (110) NdGaO$_3$ substrate. Annealing conditions: 1 h at 1000 °C in O$_2$ flow (50 sccm). (b) AFM image (0.75×0.75 μm$^2$) and the corresponding RHEED pattern of ~ 1 monolayer SrO deposited on (001) NdGaO$_3$ substrate chemically etched in m-BHF solution and then annealed for 2 h at 1000 °C in O$_2$ flow (200 sccm).

**Figure 7.** AFM images of two (110) NdGaO$_3$ substrates: (a) chemically etched for 45 s in m-BHF solution and then annealed for 1 h at 1000 °C in O$_2$ flow (100 sccm), and (b) (only) annealed for 1 h at 950 °C in O$_2$ flow (100 sccm). Both samples were stored in air (average humidity of 30 %) for 1 month. Scans size: (a) 1.5×1.5 μm$^2$, (b) 2×2 μm$^2$.

iii. Two (110) NdGaO$_3$ samples, one chemically treated and the other one only annealed, were stored in air for 1 month. As showed by the AFM observations (figure 7), the chemically treated surface was not affected by the contact with moisture from the air (see figure 7(a)). In contrast, the surface of the annealed substrate degraded in time (figure 7(b)). From the difference in stability in air of the NdO$_{1+\delta}$ and GaO$_{2-\delta}$ surfaces (i.e., while GaO$_{2-\delta}$ is stable, NdO$_{1+\delta}$ forms hydroxides and carbonates in contact with the moisture from air) [19], it may be concluded that the annealed NdGaO$_3$ substrate has a NdO$_{1+\delta}$ top layer, while the etched surface has a predominantly GaO$_{2-\delta}$ termination. This result shows that care should be taken when storing the as-received and annealed substrates for longer time in air, as this may lead to surface degradation.

## 3. Conclusions

In thin film epitaxy, the quality of the substrate surface morphology and the chemistry of the terminating layer are important parameters that determine the film growth. In this paper we described a chemical etching procedure developed to obtain GaO$_{2-\delta}$ single terminated (110) or (001) NdGaO$_3$ substrates. Modified BHF solutions, with controlled pH, were prepared and used in order to improve the surface morphology of substrates characterized by relatively low miscut angle, whereas for higher miscut angles (e.g., $\alpha > 0.3°$) a buffered HCl solution was developed with similar results. By considering the layered structure of the substrate, selective removal of one of the surface oxides (i.e., NdO$_{1+\delta}$) was achieved by this procedure. The resulting surface morphology was studied ex-situ by AFM and in-situ by high-pressure

RHEED. There is a strong dependence of the final surface morphology on the properties of the substrate (e.g., surface chemistry, miscut angle and its orientation with crystallographic axes), on the characteristics of the etching solution, and on the annealing conditions (temperature and time). Proper chemical etching associated with a subsequent thermal treatment resulted in structurally and chemically well defined surfaces, free of etch pits. For the experimental conditions described in this paper, annealing yielded a $NdO_{1+\delta}$-terminated surface, while selective chemical etching using $NH_4F$- or $NH_4Cl$-based solutions gave a predominantly $GaO_{2-\delta}$-terminated surface. Between (001) and (110) $NdGaO_3$ substrates, a lower free surface energy was observed for former, as well as for the $GdO_{2-\delta}$-terminated surface, by comparison with the $NdO_{1+\delta}$-terminated $NdGaO_3$ surface.

**Acknowledgements**

This work was financed by the Dutch Science Foundation (stichting Fundamenteel Onderzoek der Materie).